\documentclass[preprint,1p]{elsarticle}
\usepackage{amsmath,amssymb}     
\usepackage{color}              
\usepackage{cuted}
\usepackage{amssymb}

\journal{Physics Letters B}

\begin{document}
	
	\begin{frontmatter}
		
		
		
		\title{Integrable Boundary States  from Maximal Giant Gravitons in ABJM Theory}
		
		
		\author{Peihe Yang}
		
		\affiliation{organization={Center for Joint Quantum Studies and Department of Physics, School of Science, Tianjin University},
			addressline={135 Yaguan Road}, 
			city={Tianjin},
			postcode={300350}, 
			country={China}}
		
		\begin{abstract}
			We investigate the integrability of the boundary state arising from the subdeterminant operators in the alternating $SU(4)$ spin chain in ABJM theory. Our findings show that the resulting matrix product states are only integrable for two special giant gravitons, including the maximal giant graviton. Furthermore, we extend our analysis to more general boundary states.
		\end{abstract}
		
		
		
		
	\end{frontmatter}
	
	
	
	\section{Introduction}
	Integrability has been developed to study the spectrum of $\mathcal{N}$=4 supersymmetric Yang–Mills (SYM) theory, Minahan and Zarembo \cite{Minahan:2002ve} first discovered that the spectrum of the single-trace operator at one-loop order can be mapped to the spectrum of an integrable spin chain. Then integrability was used to investigate the all-loop spectrum of single-trace operator in the planar limit \cite{Beisert:2010jr,Gromov:2014caa,Gromov:2009tv, Bombardelli:2009ns,Arutyunov:2009ur,Gromov:2013pga,Gromov:2017blm}. Various developments subsequently generalized it to the computation of the correlation functions-including three-point functions and hexagon approach \cite{Basso:2015zoa,Fleury:2016ykk,Basso:2015eqa,Basso:2017muf,Eden:2015ija,Fleury:2017eph,DeLeeuw:2019dak,Escobedo:2010xs,Escobedo:2011xw,Gromov:2011jh,Gromov:2012uv,Bissi:2012ff}, four- point functions \cite{Basso:2017khq,Eden:2016xvg,Coronado:2018ypq}, octagon\cite{Belitsky:2019fan,Kostov:2019stn,Kostov:2019auq,Fleury:2020ykw,Bargheer:2019kxb,Bargheer:2019exp,Belitsky:2020qrm}, three-point functions \cite{Jiang:2019xdz,Basso:2022nny} and the exact overlap formula in spin chain \cite{Jiang:2020sdw}. Recently, the three-point functions consisting of two determinat operators and one non-BPS single trace operator were systematically studied in $\mathcal{N}=4$ SYM \cite{Jiang:2019zig,Jiang:2019xdz,Bissi:2011dc}, the idea was then generalized to ABJM theory \cite{Yang:2021hrl,Yang:2021kot,Hirano:2012vz} which is integrable in the planar limit \cite{Minahan:2008hf,Bak:2008cp, Arutyunov:2008if,Stefanski:2008ik}. For more details, please refer to the review paper \cite{Klose:2010ki}.
	
	The effective field theory developed in \cite{Jiang:2019zig,Jiang:2019xdz} helps us convert the determinant operators into a matrix product states (MPS) at weak coupling. In this paper, we focus on the scalar sector of ABJM theory which can be described as an alternating $SU(4)$ spin chain. Just like the case in $\mathcal{N}=4$ SYM, the determinant operators can be mapped to a matrix product state \cite{Yang:2021hrl,Yang:2021kot}. The three point functions can hence be calculated by the contraction of the matrix product state and the Bethe state. One of the main results of \cite{Yang:2021hrl} is that some selection rules are necessary condition of non-vainshing overlap.

	The matrix product state is not only relevant to the three-point function in ABJM theory, but also appears in the one-point function associated with a codimension one defect, such as $D3$-$D5$ defect one point functions \cite{deLeeuw:2015hxa,Buhl-Mortensen:2015gfd,Buhl-Mortensen:2016pxs,Buhl-Mortensen:2017ind,deLeeuw:2016umh,Kristjansen:2020mhn,Komatsu:2020sup}, $D3$-$D7$ defect one point functions \cite{DeLeeuw:2018cal,deLeeuw:2016ofj,DeLeeuw:2019ohp,GimenezGrau:2018jyp}, ABJM domain wall \cite{Kristjansen:2021abc}, the duality relation for the overlap formula  \cite{Kristjansen:2021xno,Kristjansen:2020vbe}. Recent works  \cite{Linardopoulos:2022wol,Linardopoulos:2021rfq} suggest that the ABJM domain wall is integrable for all loop orders and for any value of the bond dimension. In \cite{Piroli:2017sei}, the authors propose a criterion of integrable boundary state, subsequencely the nested $\emph{K}$ matrices method of overlap formula is developed in \cite{Gombor:2020kgu,Gombor:2020auk,Gombor:2021hmj}.
	We find that the matrix product state in \cite{Yang:2021hrl} is integrable according to the criteria of \cite{Piroli:2017sei}, and the non-vanishing three-point functions provide us with certain selection rules. These selection rules that are relevant to the integrable boundary are referred to as root pairing. The root pairing can therefore be derived from the integrable boundary condition. Finally, we investigate the integrability of the giant graviton in \cite{Yang:2021hrl} and find that it is only integrable for two cases. One of these cases corresponds to the maximal giant graviton in \cite{Yang:2021hrl}, while the other case has a string theory realization but does not have a field theory explanation. This result sheds light on the underlying integrable structure of giant gravitons in ABJM theory.
	
	The author has observed that \cite{Gombor:2022aqj}  proves a set of integrable matrix product states in ABJM theory and proposes an overlap formula for alternating $SU(4)$ spin chain by using the nested $K$ matrices method developed in \cite{Gombor:2020kgu,Gombor:2020auk}. The method proposed in \cite{Gombor:2021hmj} proves the integrable condition for a class of boundary states and provides a $K$ matrix proof of the integrable boundary condition. In this paper, we will present an alternative approach that directly checks the integrable boundary conditions and sequentially generalizes them to a more general matrix product state. Our method avoids the difficulty of solving the Yang-Baxter equation for the $K$ matrix. In the present paper we prove that the
	MPS obtained in \cite{Yang:2021hrl} is integrable in the sense of \cite{Piroli:2017sei}. The root pairing proposed in \cite{Yang:2021hrl} can be derived from the integrable condition and we generalize the result to a more general matrix product state. In section 2, we first review some results in \cite{Yang:2021hrl}, including the algebraic Bethe ansatz method and the selection rules. In section 3, we give the integrable condition for $SU(4)$ alternating spin chain and propose a method to prove it. In section 4, we find the previous method can be generalized to a more general state. In section 5, we prove the giant graviton \cite{Yang:2021hrl} is integrable for the cases where $\omega=0$ or 1. Finally, we conclude in section 6.

	\section{Selection Rules}
	
	Let us first review the basic fact about the integrability of ABJM theory at two loop order, the two-loop dilatation operators in the scalar sector of ABJM theory can be described as the $SU(4)$ alternating spin chain. The Hamiltonian is given by 
	\begin{align}
		H=\dfrac{\lambda^2}{2}\sum_{l=1}^{2L}(2-2 P_{l,l+2}+P_{l,l+2}K_{l,l+1}+K_{l,l+1}P_{l,l+2}).
	\end{align}
	The odd/even sites of the alternativing spin chain sit in the fundamental/antifundamental representation.
	\begin{align}
		|Y^{A_1}\bar{Y}_{A_2}Y^{A_3}\bar{Y}_{A_4}\cdots\rangle
	\end{align}
	
	The $SU(4)$ alternating chain has four $\emph{R}$ matrices,
	\begin{align}
		& R_{ab}=u \mathrm{I}+P_{ab}\\\nonumber
		& R_{\bar{a}\bar{b}}=u\mathrm{I}+P_{\bar{a}\bar{b}}\\\nonumber
		& R_{a \bar{b}}=-(u+2)\mathrm{I}+K_{a \bar{b}}\\\nonumber
		& R_{\bar{a}b}=-(u+2)\mathrm{I}+K_{\bar{a}b},
	\end{align} 
	where a and $\bar{a}$ denote the fundamental/anti-fundamental representation of $SU(4)$. We can define two different monodromy matrices by introducing two different reprenstations of auxiliary spaces.
	\begin{align}
		&T_a(u)=R_{a1}(u)R_{a\bar{1}}(u)\cdots R_{aL}(u)R_{a\bar{L}}(u)\\\nonumber
		&T_{\bar{a}}(u)=R_{\bar{a}1}(u)R_{\bar{a}\bar{1}}(u)\cdots R_{\bar{a}L}(u)R_{\bar{a}\bar{L}}(u).
	\end{align}
	Taking the trace of the auxiliary spaces gives the transfer matrices
	\begin{align}
		\tau(u)=\mathrm{tr}_a T_a(u), \quad \bar{\tau}(u)=\mathrm{tr}_{\bar{a}} T_{\bar{a}}(u).
	\end{align}
	Two Hamiltonians can be generated from the transfer matrices
	\begin{align}
		H_1=(\tau(0))^{-1} \dfrac{d}{du}\tau(u)|_{u=0},\quad H_2=(\bar{\tau}(0))^{-1} \dfrac{d}{du}\bar{\tau}(u)|_{u=0}.
	\end{align}
	Then the two-loop anomalous dimension matrix of scalar sector of ABJM theory can be mapped to the Hamiltonian $H=H_1+H_2$ (up to the rescale of the Hamiltonian and the a constant shift).
	
	It is useful to write down the indices of the operators
	\begin{align}
		I^{a A}_{b B}=\delta^a_b \delta^A_B,\quad P^{a A}_{b B}=\delta^a_B \delta^A_b,\quad K^{a A}_{b B}=\delta^{aA} \delta_{b B},
	\end{align}
	where index a denotes the index of the auxiliary space and $A$ acts on the Hilbert space of the spin chain. Then the transfer matrix $\tau(u)$ can be written as
	\begin{align}
		\tau(u)=(u I+P)^{a_1 A_1}_{a_2 B_1}(-(u+2)+K)^{a_2 \bar{A}_2}_{a_3 \bar{B}_2}\cdots (u I+P)^{a_{2L-1}A_{2L-1}}_{a_{2L}B_{2L-1}}(-(u+2)+K)^{a_{2L} \bar{A}_{2L}}_{a_1 \bar{B}_{2L}}.
	\end{align}
	The ground state in spin chain language is 
	\begin{align}
		\mathrm{tr}(Y^1\bar{Y}_4)^L\leftrightarrow |\underbrace{1\bar{4}\cdots 1\bar{4}}_{\text{Length 2L}}\rangle.
	\end{align}
	A general state is characterized by three sets of Bethe roots $\mathbf{u,v,w}$, where these rapidities satisfy the Bethe equation
	\begin{align}
		&1=\left(\dfrac{u_j+\frac{i}{2}}{u_j-\frac{i}{2}}\right)^L \prod_{\substack{k=1\\k\neq j}}^{K_\mathrm{u}} S(u_j,u_k) \prod_{\substack{k=1}}^{K_\mathrm{w}} \tilde{S}(u_j,w_k) \\
		&1=\prod_{\substack{k=1\\k\neq j}}^{K_\mathrm{w}} S(w_j,w_k) \prod_{\substack{k=1}}^{K_\mathrm{u}} \tilde{S}(w_j,u_k)\prod_{\substack{k=1}}^{K_\mathrm{v}} \tilde{S}(w_j,v_k)\\
		&1=\left(\dfrac{v_j+\frac{i}{2}}{v_j-\frac{i}{2}}\right)^L \prod_{\substack{k=1\\k\neq j}}^{K_\mathrm{v}} S(v_j,v_k) \prod_{\substack{k=1}}^{K_\mathrm{w}} \tilde{S}(v_j,w_k),
	\end{align}
	where $K_\mathrm{u}, K_\mathrm{v}, K_\mathrm{w}$ denote the number of $\mathrm{u}, \mathrm{v}, \mathrm{w}$, and $\mathbf{u}=\{u_1,\cdots, u_{K_u}\}, \mathbf{v}=\{v_1,\cdots, v_{K_v}\},$  $\mathbf{w}=\{w_1,\cdots, w_{K_w}\}$ and 
	\begin{align}
		S(u,v)=\dfrac{u-v-i}{u-v+i},\quad \tilde{S}(u,v)=\dfrac{u-v+\frac{i}{2}}{u-v-\frac{i}{2}}.
	\end{align}
	
	The eigenvalues of the transfer matrices can be obtained from the nested Bethe ansatz.
	\begin{align}\label{eigenvalue}
		\Lambda(u)=&(-u-2)^L (u+1)^L \prod_{i=1}^{K_u}\dfrac{u-iu_i-1/2}{u-iu_i+1/2}\\\nonumber
		&+(-u)^L(u+1)^L\prod_{i=1}^{K_v}\dfrac{u-iv_i+5/2}{u-iv_i+3/2}\\\nonumber
		&+(-u)^L(u+2)^L\prod_{i=1}^{K_u}\dfrac{u-iu_i+3/2}{u-iu_i+1/2}\prod_{i=1}^{K_w}\dfrac{u-iw_i}{u-iw_i+1}\\\nonumber
		&+(-u)^L(u+2)^L\prod_{i=1}^{K_v}\dfrac{u-iv_i+1/2}{u-iv_i+3/2}\prod_{i=1}^{K_w}\dfrac{u-iw_i+2}{u-iw_i+1},
	\end{align}
	\begin{align}
		\bar{\Lambda}(u)=& (-u)^L(u+1)^L\prod_{i=1}^{K_u}\dfrac{u-iu_i+5/2}{u-iu_i+3/2}\\\nonumber
		&+(-u-2)^L (u+1)^L\prod_{i=1}^{K_v}\dfrac{u-iv_i-1/2}{u-iv_i+1/2}\\\nonumber
		&+(-u)^L(u+2)^L\prod_{i=1}^{K_u}\dfrac{u-iu_i+1/2}{u-iu_i+3/2}\prod_{i=1}^{K_w}\dfrac{u-iw_i+2}{u-iw_i+1}\\\nonumber
		&+(-u)^L(u+2)^L\prod_{i=1}^{K_v}\dfrac{u-iv_i+3/2}{u-iv_i+1/2}\prod_{i=1}^{K_w}\dfrac{u-iw_i}{u-iw_i+1}.
	\end{align}
	
	Here we use the convention of the appendix D of \cite{Yang:2021hrl} where we propose a conjecture about the overlap of the three point functions. The three point fuctions  consisting of two maximal giant gravitons and one non-BPS operator can be written as the overlap between a boundary state $\langle \mathcal{B}|$ and the $SU(4)$ Bethe state $|\mathbf{u,v,w}\rangle$ in the planar limit. For convenience, we present the formula of  $\langle \mathcal{B}|$
	\begin{align}\label{Booundary state}
		\langle \mathcal{B}|=\sum_{B_s,\bar{B}_s=1,4}\langle B_1\bar{B}_1\cdots B_L\bar{B}_L|(1+(-1)^J) ,
	\end{align}
	where $\emph{J}$ is the $U(1)$ charge of the boundary state $J=L-(\text{number of 4 on odd sites})-(\text{number of $\bar{1}$ on even sites})$.
	
	The single-trace operator exhibits the cyclicity property, leading to the zero-momentum condition within the spin chain framework.
	\begin{align}
		1=\prod_{j=1}^{K_u}\frac{u_j+i/2}{u_j-i/2}\prod_{j=1}^{K_v}\frac{v_j+i/2}{v_j-i/2}
	\end{align}
	
	The overlaps are non-vanishing only when they satisfy specific selection rules, which have been confirmed through numerical checks \cite{Yang:2021hrl}.
	
	0. The boundary state only contains $Y^1,Y^4,\bar{Y}_1,\bar{Y}_4$. In the coordinate Bethe ansatz side, generating such states requires $K_u=K_v=K_w$.
	
	1. The U(1) charge $\emph{J}$ of the state is even.
	
	2. \textbf{Root pairing}: The rapidities of the Bethe roots satisfy $ \textbf{u}=-\textbf{v}, \textbf{w}=-\textbf{w}$.
	
	The first selection rule can directly derived from the formula of the boundary state.
	
	In this note, we will prove that our boundary state \eqref{Booundary state} satisfies the integrable boundary condition which leads to the root pairing.
	\section{Integrable Boundary State}
	We can remove the factor $1+(-1)^J$ in the boundary state if we focus on the Bethe state that satisfy the selection rule 1. Let $|\Psi\rangle$ denote the boundary state $\sum_{B_s,\bar{B}_s=1,4}\langle B_1\bar{B}_1$\\$\cdots B_L\bar{B}_L|$. Later we will prove that our boundary state $|\Psi\rangle$ satisfies the following integrable condition
	\begin{align}\label{boundary int}
		\Pi \tau(u)\Pi|\Psi\rangle=\tau(u)|\Psi\rangle,
	\end{align}
	where $\Pi$ is the parity operator		
	\begin{align}
		\Pi|B_1\bar{B}_1\cdots B_L\bar{B}_L\rangle=|\bar{B}_LB_L\cdots \bar{B}_1B_1\rangle.
	\end{align}
	The integrable boundary condition is firstly proposed in \cite{Piroli:2017sei} by requiring the state to be annihilated by the odd conserved charges. \eqref{boundary int} implies the odd conserved charge annilate the boundary state. The authors of \cite{Gombor:2020kgu} then generalize it to the chiral/achiral condition to distinguish the pair condition. Here chiral pair structure only contains $\mathbf{u}=-\mathbf{u}, \mathbf{v}=-\mathbf{v}, \mathbf{w}=-\mathbf{w}$ and achiral pair condition contains $\mathbf{u}=-\mathbf{v}, \mathbf{w}=-\mathbf{w}$. 
	
	The left hand side (LHS) is equivalent to
	\begin{align}
		\Pi \mathrm{tr}_{a}(R_{a 1}(u)R_{a \bar{1}}(u)\cdots R_{a L}(u)R_{a  \bar{L}}(u))|\bar{B}_LB_L\cdots \bar{B}_1B_1\rangle.
	\end{align}
	The crossing symmetry of the $\emph{R}$ matrices relates the transpose of the auxiliary space to the anti-fundamental R matrix.
	\begin{align}
		R_{\bar{a}i}(u)=R^{t_0}_{ai}(-u-2).
	\end{align} 
	where $t_0$ is the transpositon of the auxiliary space.
	
	Transposing the auxiliary space index will reverse the order of $\emph{R}$ matrices in auxiliary space,
	\begin{align}
		&\Pi \mathrm{tr}_{\bar{a}}(R_{\bar{a} \bar{L}}(-u-2)R_{\bar{a} L}(-u-2)\cdots R_{\bar{a} \bar{1}}(-u-2)R_{\bar{a}1}(-u-2))|\bar{B}_LB_L\cdots \bar{B}_1B_1\rangle\\\nonumber&=\mathrm{tr}_{a}(R_{a 1}(-u-2)R_{a \bar{1}}(-u-2)\cdots R_{a L}(-u-2)R_{a \bar{L}}(-u-2))|B_1\bar{B}_1\cdots B_L\bar{B}_L\rangle.
	\end{align}
	
	The integrable boudary condition becomes 
	\begin{align}\label{integrbale condition}
		\tau(-u-2)|\Psi\rangle=\tau(u)|\Psi\rangle.
	\end{align}
	The overlap between the boundary state and the Bethe state gives
	\begin{align}
		\langle \Psi|\tau(u)|\textbf{u,v,w}\rangle=\langle \Psi|\Pi\bar{\tau}(u)\Pi|\textbf{u,v,w}\rangle=\langle \Psi|\tau(-u-2)|\textbf{u,v,w}\rangle.
	\end{align}
	
	The second pair condition can be obtained by substituting the eigenvalue of the transfer matrix
	\begin{align}
		\Lambda(u)=\Lambda(-u-2)\Rightarrow\textbf{u}=-\textbf{v}, \textbf{w}=-\textbf{w}.
	\end{align}
	The rest of this note is the proof of \eqref{integrbale condition}.
	\subsection{Proof of \eqref{integrbale condition}}
	\eqref{integrbale condition} can be written as 
	\begin{align}\label{identity 0}
		&\sum_{\bar{B}_i,B_i=1,4}\mathrm{tr}[(u I+P)(-(u+2)I+K )\cdots] |B_1\bar{B}_1\cdots\rangle=\\\nonumber &\sum_{\bar{B}_i,B_i=1,4} \mathrm{tr}[(-(u+2)I+P)(u I+K)\cdots] |B_1\bar{B}_1\cdots\rangle.
	\end{align}
	The proof of \eqref{identity 0} can be summarized as:

	\textbf{Main idea}: By expanding the transfer matrices on both sides of \eqref{identity 0}, we obtain various terms with different coefficients. We then classify these terms into different classes. For each class on the left-hand side (LHS), we can assign the corresponding class on the right-hand side (RHS) with the same coefficients. This assignment ensures that when these classes act on our boundary state, they yield the same result. Then we complete the proof.
	
	Now, let us provide a detailed proof of equation \eqref{identity 0}.
	
	The main difference between the LHS and RHS of equation \eqref{identity 0} lies in the different coefficients of the identity operator on the odd and even sites. We can decompose the transfer matrices based on these coefficients.
	
	It is convenient to give the explicit form of the contractions of $P,K,I$
	\begin{align}\label{identity1}
		&(K)^{a \bar{A}_1}_{b \bar{B}_1} (P)^{b A_2}_{c B_2}=\delta^{a \bar{A}_1}\delta_{b \bar{B}_1}\delta^{b}_{B_2}\delta_{c}^{A_2}=\delta_{\bar{B}_1,B_2}\delta^{a \bar{A}_1}\delta_{c}^{A_2}\\
		&(P)^{a A_1}_{bB_1} (K)^{b \bar{A}_2}_{e \bar{B}_2}=\delta^a_{B_1}\delta^{b \bar{A}_2}\delta^b_{A_1}\delta_{e \bar{B}_2}=\delta^{a}_{B_1}\delta^{A_1 \bar{A}_2}\delta_{e \bar{B}_2}\\
		&(K)^{a \bar{A}_1}_{b\bar{B}_1}  (K)^{b \bar{A}_2}_{c \bar{B}_2}=\delta^{a\bar{A}_1}\delta_{\bar{B}_1b}\delta^{\bar{A}_2b}\delta_{c\bar{B}_2}=\delta^{a\bar{A}_1}\delta^{\bar{A}_2}_{\bar{B}_1}\delta_{c\bar{B}_2}\\
		&(P)^{a A_1}_{bB_1}  (P)^{b A_2}_{c B_2}=\delta^{a}_{B_1}\delta_{c}^{A_2}\delta_{b}^{ A_1}\delta^{b}_{B_2}=\delta^{a}_{B_1}\delta_{c}^{A_2}\delta_{B_2}^{A_1}.
	\end{align}
	Here, $a$ and $b$ represent the indices of the auxiliary space, while $A$ and $B$ denote the actions on the odd sites of the Hilbert space of the spin chain, and $\bar{A}$ and $\bar{B}$ represent the actions on the even sites of the Hilbert space of the spin chain.
	
	A key observation is that the coefficients in front of $\mathrm{tr}(PI\cdots IK)$ on both sides of \eqref{identity 0} are
	\begin{align}
		\text{coefficients of}\quad \mathrm{tr}(PI\cdots IK)=(-u^2-2u)^{\text{(number of I)}/2}.
	\end{align}
	Here $\cdots$ signifies that only the identity operator exists between the $P$ and $K$ operators.
	
	The expansion of transfer matrices can be classified based on whether they contain $P I\cdots IK$ or not
	\begin{align}\label{transfer}
		\text{transfer matrix}&=\text{terms containing $P I\cdots IK$}+\text{terms not containing $P I\cdots IK$}.
	\end{align}
	
	The first part in \eqref{transfer} can be further decomposed 
	\begin{align}\label{class1}
		\text{terms without $P I\cdots IK$}= \mathrm{tr}(IIII\cdots IIII)+\text{terms contain only P or K}.
	\end{align}
	The matching between terms that only contain $I$ is straightforward. Now, let's shift our focus to the second part of equation \eqref{class1}.
	
	The condition that the transfer matrix does not contain $PI\cdots IK$ implies that these terms only contain $P$ or $K$ in the expansion of the transfer matrices. Now, we claim that we have the following correspondence between the LHS and the RHS of equation \eqref{identity 0}. 
	\begin{align}\label{proof0}
		&\mathrm{tr}(\cdots\underbrace{P_lI_l}_{l\text{th factor}}\cdots\underbrace{P_iI_i}_{i\text{th factor}}\cdots \underbrace{P_jI_j}_{j\text{th factor }}\cdots)|\Psi\rangle \quad\text{in LHS/RHS}=\\\nonumber &\mathrm{tr}(\cdots\underbrace{I_lK_l}_{l\text{th factor}}\cdots\underbrace{I_iK_i}_{i\text{th factor}}\cdots \underbrace{I_j K_j}_{j\text{th factor }}\cdots)|\Psi\rangle \quad\text{in RHS/LHS}.
	\end{align}
	\eqref{proof0} requires that the appearance of the $P$ operator on one side should correspond to the $K$ operator on the other side at the same position.
	
	The coefficients match precisely because the coefficient of $PI/IK$ in LHS is equal to the coefficient of $IK/PI$ in RHS. For the case only contains one P or K, the transfer matrix has the corresponding index $\delta_{A_iB_i}$ or $ \delta_{\bar{A}_i\bar{B}_i}$. Then we have
	\begin{align}
		\mathrm{tr}(\cdots\underbrace{P_iI_i}_{i\text{th factor}}\cdots)|\Psi\rangle \quad\text{in LHS/RHS}=\mathrm{tr}(\cdots\underbrace{I_iK_i}_{i\text{th factor}}\cdots)|\Psi\rangle \quad\text{in RHS/LHS}.
	\end{align}
	
	For cases involving more than two $P/K$s, the contraction gives the Kronecker delta function $\delta_{A_i B_j}/\delta_{\bar{A}_i\bar{B}_i}$ which induces the permutation between indices on odd/even site.
	For example,
	\begin{align}
		\mathrm{tr}(PIPI)|\Psi\rangle=\sum_{B_i,\bar{B}_i=1,4}|B_2\bar{B}_1 B_1\bar{B}_2\rangle.
	\end{align}
	$\mathrm{tr}(IKIK)|\Psi\rangle$ has the similar result $\sum_{B_i,\bar{B}_i=1,4}|B_1\bar{B}_2 B_2\bar{B}_1\rangle$. 
	
	It can be directly generalized to the length 2L spin chain, PP contraction induces the permutation on the odd sites while KK contraction induces the permutation on the even sites. Our boundary state is invariant under the permutation at the odd/even sites. i.e.
	\begin{align}
		\mathrm{tr}(\cdots\underbrace{ P_iI_i}_{i\text{th factor}}\cdots \underbrace{P_jI_j}_{j\text{th factor}}\cdots)|\Psi\rangle =\mathrm{tr}(\cdots\underbrace{I_iK_i}_{i\text{th factor}}\cdots \underbrace{I_j K_j}_{j\text{th factor}}\cdots)|\Psi\rangle=\sum_{B_i,\bar{B}_i\cdots=1,4}|B_1\bar{B}_1\cdots\rangle.
	\end{align}
	
	The next step is to establish the correspondence between the terms that contain the $PI\cdots IK$ factor in \eqref{transfer}.
	
	According to the previous argument, we can always divide the entire spin chain into $P I\cdots I K$ factors and the interval between the $PK$ factors. The operators in the interval between two PK factor are $IK\cdots IK\cdots PI\cdots PI$. Hence the most general form of the building block is 
	\begin{align}\label{identity2}
		\underbrace{PI\cdots IK}_{\text{PK factor}} II\cdots \underbrace{IK\cdots IK\cdots}_{\text{number of K}=m}II\cdots \underbrace{PI\cdots PI\cdots}_{\text{number of P}=n} II\cdots\underbrace{PI\cdots IK}_{\text{PK factor}}.
	\end{align}
	Here, we insert some $K$ operators and $P$ operators in the interval between two $PK$ factors. It is important to note that the $K$ operators should be on the left-hand side of the $P$ operators within the interval. Hence, the entire transfer matrices are divided into different regions that are separated by the $PK$ factors.
	
	We claim the following correspondence in \eqref{identity 0}
	\begin{align}\label{proof1}
		&&\mathrm{tr}(\cdots\underbrace{{\color{red}PI\cdots IK}}_{\text{PK factor}}\underbrace{II\cdots}_{\text{only I}} \underbrace{IK\cdots IK\cdots}_{\text{number of K}=m} \underbrace{II\cdots}_{\text{only I}}\underbrace{PI\cdots PI\cdots}_{\text{number of P}=n} \underbrace{II\cdots}_{\text{only I}}\underbrace{ {\color{magenta}PI\cdots IK}}_{\text{PK factor}}\cdots)|\Psi\rangle\leftrightarrow\\\nonumber &&\mathrm{tr}(\cdots\underbrace{{\color{red}PI\cdots IK}}_{\text{PK factor}} \underbrace{II\cdots}_{\text{only I}}\underbrace{IK\cdots IK\cdots}_{\text{number of K}=n}\underbrace{II\cdots}_{\text{only I}} \underbrace{PI\cdots PI\cdots}_{\text{number of P}=m} \underbrace{II\cdots}_{\text{only I}}\underbrace{{\color{magenta}PI\cdots IK}}_{\text{PK factor}}\cdots)|\Psi\rangle.
	\end{align}
	Now the general form contains three parts: PK factor, IK factors and PI factors in the interval. Here the $PK$ factors have the same color to indicate that they are on the same site on the spin chain. The coefficient of $PI,IK$ is $-(u+2),u$ at LHS and $u,-(u+2)$ at RHS, so we require the number of $IK,PI$ at LHS is equal to the number of $PI,IK$ at RHS. The sites of the $PI, IK$ factors in the interval of LHS/RHS of \eqref{proof1} are labeled by $\mathcal{C}/\mathcal{D}$. We require $\mathcal{C}=\mathcal{D}$. For example,
	\begin{align}
		&&\mathrm{tr}(\underbrace{{\color{red}PIIK}}_{\text{PK factor }}\underbrace{{\color{blue}IK_i}\cdots}_{\text{number of K}=1} \underbrace{{\color{green}PI_j}\cdots {\color{cyan}PI_k}\cdots {\color{black}PI_l}}_{\text{number of P}=3} \underbrace{ {\color{magenta}PI IK}}_{\text{PK factor}})|\Psi\rangle \leftrightarrow\\\nonumber &&\mathrm{tr}(\underbrace{{\color{red}PI IK}}_{\text{PK factor}}\underbrace{{\color{blue}IK_i}\cdots {\color{green}IK_j}\cdots {\color{cyan}IK_k}}_{\text{number of K}=3} \underbrace{{\color{black}PI_l}\cdots}_{\text{number of P}=1} \underbrace{{\color{magenta}PIIK}}_{\text{PK factor}})|\Psi\rangle.
	\end{align}
	Here the same color means the same site on the spin chain and the $IK_i, PI_j$ are marked by their sites on the spin chain, hence we have $\mathcal{C}=\mathcal{D}=\{i,j,k,l\}$. Here we have just presented a building block of the transfer matrices, the above correspondence can be generalized to the entire transfer matrices.
	
	In general, there are three types of Kronecker delta function:\eqref{identity2}, $\delta_{A_i A_j}$ comes from the PK factor (PK contraction),  $\delta_{B_o B_p}$ comes from the $KP$ contraction,  $\delta_{A_i B_j}$ comes from the KK and PP contraction in the interval. That is
	\begin{align}\label{identity3}
		\underbrace{\delta^{A_i \bar{A}_j}\delta^{A_k \bar{A}_L}\cdots}_{\text{m PK factors}}\underbrace{\delta_{B_m \bar{B}_n}\delta_{B_o \bar{B}_p}\cdots}_{\text{n KP contractions}}\underbrace{\delta^{A_m}_{B_z}\delta^{ \bar{A}_n}_{\bar{B}_p}\cdots}_{\text{KK and PP contractions}}.
	\end{align}
	Note that the number of A indices should be equal to the number of B indices, so we conclude m=n.
	
	$\delta^{A_i}_{B_j}$: The arugment in case 1 suggests that $\delta^{A_i}_{B_j}$ comes from $PP\cdots P$ contraction permutes the orginial index $B$ on odd site and $\delta^{ \bar{A}_i}_{\bar{B}_j}$ comes from $K\cdots KK$ contraction permutes the orginial index $B$ on even site. Summing over these sites gives the same result.
	
	$\delta_{B_o \bar{B}_p}$: Summing over the original index B gives us a const,
	\begin{align}
		\sum_{\bar{B}_i, B_i=1,4} \delta_{B_o \bar{B}_p}\delta_{B_m \bar{B}_m}\cdots=2^n,
	\end{align}
	where n is the number of $IKI\cdots IPI$ contraction.
	
	$\delta^{A_i \bar{A}_j}$: $\delta^{A_i \bar{A}_j}$ denote the sum of the final state index,
	\begin{align}
		\delta^{A_i \bar{A}_j}\leftrightarrow \sum_{\tilde{A}=1,2,3,4} |\cdots \underbrace{\tilde{A}}_{i-th site}\cdots \underbrace{\tilde{A}}_{j-th site}\rangle.
	\end{align}
	Both sides of \eqref{proof1} will give the same result once we fix the position of the PK factor.
	
	Hence, both sides of \eqref{proof1} are
	\begin{align}
		2^n \sum_{B_i,\bar{B}_i\cdots=1,4\atop \tilde{A}_i=1,2,3,4}|\cdots \tilde{A}_1\bar{B}_i B_j\tilde{A}_1 B_k\bar{B}_l\cdots B_m\bar{B}_n \tilde{A}_2\bar{B}_o A_p \tilde{A}_2\cdots \tilde{A}_3\bar{B}_rB_s\tilde{A}_3\cdots\rangle.
	\end{align}
	
	For the case that only contains one $PK$ factor, the argument is similar. By applying equations \eqref{proof0} and \eqref{proof1} to every $PK$ factor and its corresponding interval, we can establish the desired correspondence expressed in equation \eqref{identity 0}.
	
	\section{More General State}
	The method in previous section can be generalized to a special matrix product state. That is,
	\begin{align}\label{state}
		|\Psi^\prime\rangle =\sum_{B_i,\bar{B}_i=1,2,3,4}\mathrm{tr}(t_{B_1}\bar{t}_{\bar{B}_1}\cdots t_{B_2}\bar{t}_{\bar{B}_2}\cdots) |B_1\bar{B}_1\cdots B_2\bar{B}_2\cdots\rangle,
	\end{align}
	where $t_i,\bar{t}_i$ commute to each other. The previous argument suggests that the transfer matrix can be split as \eqref{transfer} and we will prove the integrable boundary condition in two steps.
	
	1. For the part of the transfer matrix that does not contain $PK$ factors, we need to prove 
	\begin{align}\label{new1}
		\mathrm{tr}(\cdots P_i I_i\cdots)|\Psi^\prime\rangle\leftrightarrow \mathrm{tr}(\cdots  I_i K_i\cdots)|\Psi^\prime\rangle.
	\end{align}
	
	The case only contains one PI or IK is obvious. As we discussed before, the indices involving $\delta_{A_i B_j}/\delta_{\bar{A}_i \bar{B}_j}$ will permute the original index on the odd/even site. For example,
	\begin{align}
		&\sum_{B_i,\bar{B}_i}\mathrm{tr}(t_{B_1}\bar{t}_{\bar{B}_1}t_{B_2}\bar{t}_{\bar{B}_2})\mathrm{tr}(PIPI)|B_1\bar{B}_1B_2\bar{B}_2\rangle= \sum_{B_i,\bar{B}_i}\mathrm{tr}(t_{B_1}\bar{t}_{\bar{B}_1}t_{B_2}\bar{t}_{\bar{B}_2})|B_2\bar{B}_1B_1\bar{B}_2\rangle\\\nonumber
		&\sum_{B_i,\bar{B}_i}\mathrm{tr}(t_{B_1}\bar{t}_{\bar{B}_1}t_{B_2}\bar{t}_{\bar{B}_2})\mathrm{tr}(IKIK)|B_1\bar{B}_1B_2\bar{B}_2\rangle= \sum_{B_i,\bar{B}_i}\mathrm{tr}(t_{B_1}\bar{t}_{\bar{B}_1}t_{B_2}\bar{t}_{\bar{B}_2})|B_1\bar{B}_2B_2\bar{B}_1\rangle,
	\end{align}
	where $i,j,k,l\in \{1,2,3,4\}$. Rearranging the matrices $t_i,\bar{t}_j$, we obtain $\sum_{B_i,\bar{B}_i}\mathrm{tr}(t_{B_i}\bar{t}_{\bar{B}_j}t_{B_k}\bar{t}_{\bar{B}_l})$\\$|B_i\bar{B}_jB_k\bar{B}_l\rangle$.
	
	The similar logic can be directly generalized to the length L spin chain. The permutation induced by $\delta_{A_i B_j},\delta_{\bar{A}_i \bar{B}_j}$ can be canceled by the rearrangement of the matrices $t,\bar{t}$. The final state is equal to
	\begin{align}
		\sum_{B_i,\bar{B}_j}\mathrm{tr}(t_{B_i}\bar{t}_{\bar{B}_j}\cdots t_{B_k}\bar{t}_{\bar{B}_l}\cdots) |B_i\bar{B}_j\cdots B_k\bar{B}_l\cdots\rangle.
	\end{align}
	Hence,
	\begin{align}
		\mathrm{tr}(\cdots\underbrace{P_iI_i}_{i\text{th factor}}\cdots \underbrace{P_jI_j}_{j\text{th factor}}\cdots\underbrace{P_lI_l}_{\text{lth factor }}\cdots)|\Psi^\prime\rangle=\mathrm{tr}(\cdots\underbrace{I_iK_i}_{i\text{th factor}}\cdots \underbrace{I_j K_j}_{j\text{th factor}}\cdots\underbrace{I_l K_l}_{\text{lth factor }}\cdots)|\Psi^\prime\rangle.
	\end{align}
	
	2. Consider the following parts of the transfer matrix
	\begin{align}
		\underbrace{PI\cdots IK}_{\text{PK factor}} \underbrace{IKIKIK\cdots}_{\text{number of P}=m} \underbrace{PIPI\cdots}_{\text{number of K}=n} \underbrace{PI\cdots IK}_{\text{PK factor}}.
	\end{align}
	Now we want to prove the correspondence \eqref{proof1} for boundary state $|\Psi^\prime\rangle$. For convenience, we list the indices
	\begin{align}
		\underbrace{\delta^{A_i \bar{A}_j}\delta^{A_k \bar{A}_L}\cdots}_{\text{m PK factors}}\underbrace{\delta_{B_m \bar{B}_n}\delta_{B_o \bar{B}_p}\cdots}_{\text{m KP contractions}}\underbrace{\delta^{A_m}_{B_z}\delta^{ \bar{A}_n}_{\bar{B}_p}\cdots}_{\text{KK and PP contractions}}.
	\end{align}
	
	Similar to the previous case, we still encounter three types of Kronecker delta functions.
	
	$\delta^{A_i}_{B_j}$: $\delta^{A_i}_{B_j}$ permutes the state indices on the odd/even sites and can be cancelled by rearranging the matrices $t_i$ and $\bar{t}_j$.
	
	$\delta_{B_o \bar{B}_p}$: $\delta_{B_o \bar{B}p}$ restricts the matrices $\sum_{o,p}t_o\bar{t}_p$ to be $\sum_it_i\bar{t}_i$.
	
	$\delta^{A_i \bar{A}_j}$: The presence of $\delta^{A_i \bar{A}_j}$ requires us to sum over the resulting state indices.
	\begin{align}
		\delta^{A_i \bar{A}_j}\leftrightarrow \sum_{\tilde{A}_1=1,2,3,4} |\cdots\underbrace{\tilde{A}_1}_{\text{i site}}\cdots\underbrace{\tilde{A}_1}_{\text{j site}}\cdots\rangle.
	\end{align}
	
	Let us denote the indices in $\delta_{B\bar{B}}$ as $\mathcal{B}_1$ and the remaining indices as $\mathcal{B}_2$. $\delta_{B_o \bar{B}_p}$ gives coefficients
	\begin{align}
		\sum_{\mathcal{B}_1,\mathcal{B}_2}\underbrace{\delta_{B_i\bar{B}_j}\delta_{B_l\bar{B}_k}\cdots}_{\mathcal{B}_1\, \text{indices}} \mathrm{tr}({\color{red}{t}}\bar{t}_{\bar{B}_j}\cdots t_{B_i}{\color{red}{\bar{t}}}\cdots {\color{red}{t}}\bar{t}_{\bar{B}_k}\cdots t_{B_l}{\color{red}{\bar{t}}}\cdots)=\sum_{i,j\cdots,\mathcal{B}_2} \mathrm{tr}({\color{red}{t}}\bar{t}_i\cdots t_i{\color{red}{\bar{t}}}\cdots {\color{red}{t}}\bar{t}_j\cdots t_j{\color{red}{\bar{t}}}\cdots),
	\end{align}
	where the matrices with $\mathcal{B}_2$ indices are highlighted in red.
	
	So the final state is 
	\begin{align}
		\sum_{\tilde{A}_i,i\cdots,\mathcal{B}_2} \mathrm{tr}({\color{red}{t}}\bar{t}_i\cdots t_i{\color{red}{\bar{t}}}\cdots {\color{red}{t}}\bar{t}_j\cdots t_j{\color{red}{\bar{t}}}\cdots)|\tilde{A}_1{\color{red}\bar{B}B} \tilde{A}_1 \cdots \tilde{A}_2{\color{red}\bar{B}B} \tilde{A}_2\cdots\rangle,
	\end{align}
	where $\mathcal{B}_2$ indices and corresponding matrices are highlighted  in red. We find the final state does not depend on $\mathcal{B}_1$ and is invariant under the permutation of the $\mathcal{B}_2$ indices. So we complete the proof of \eqref{proof1}. Extending to the entire chain, we have the correspondence 
	\begin{align}
		\tau(u)|\Psi^\prime\rangle=\tau(-u-2)|\Psi^\prime\rangle.
	\end{align}
	The integrable boundary condition constrains the overlap
	\begin{align}
		\langle \Psi^\prime|\Lambda(-u-2)|\text{Bethe State}\rangle=\langle \Psi^\prime|\Lambda(u)|\text{Bethe State}\rangle.
	\end{align}
	$\Lambda(-u-2)=\Lambda(u)$ gives the root pairing for this MPS.
	\section{Giant graviton in ABJM theory}
	An important application of the proof presented in the previous section is to  investigate the integrable property of giant gravitons. Effective field theory reduces the three-point functions to the following boundary state \cite{Yang:2021hrl}:
	\begin{align}\label{giantgraviton}
		\mathcal{D}_{M|\mathcal{O}}=\frac{(-1)^{J+1}}{2^{\Delta-J}\sqrt{L\langle \mathcal{O}|\mathcal{O}\rangle}}\int_{0}^{2\pi}\frac{d\theta}{2\pi}\langle B_\theta|\mathcal{O}\rangle,
	\end{align} 
	with the matrix product state 
	\begin{align}
		\langle B_\theta|=\sum_{A_s,B_s=1,4}\langle A_1\bar{B}_1\cdots A_L\bar{B}_L|tr[t^{A_1}\bar{t}_{B_1}\cdots t^{A_L}\bar{t}_{B_L}],
	\end{align}
	where
	\begin{align}
		\begin{aligned}
			\mathrm{t}^{1} & =\left(\begin{array}{cc}
				e^{i \theta} \sqrt{1-\omega} & -i \sqrt{\omega} \\
				i \sqrt{\omega} & -e^{-i \theta} \sqrt{1-\omega}
			\end{array}\right), \quad \overline{\mathrm{t}}_{4}=\left(\begin{array}{cc}
				e^{i \theta} \sqrt{1-\omega} & \sqrt{\omega} \\
				\sqrt{\omega} & -e^{-i \theta} \sqrt{1-\omega}
			\end{array}\right), \\
			\mathrm{t}^{4} & =\left(\begin{array}{cc}
				-e^{i \theta} \sqrt{1-\omega} & i \sqrt{\omega} \\
				i \sqrt{\omega} & -e^{-i \theta} \sqrt{1-\omega}
			\end{array}\right), \quad \overline{\mathrm{t}}_{1}=\left(\begin{array}{cc}
				e^{i \theta} \sqrt{1-\omega} & \sqrt{\omega} \\
				-\sqrt{\omega} & e^{-i \theta} \sqrt{1-\omega}
			\end{array}\right).
		\end{aligned}
	\end{align}
	We can obtain the $K$ matrix from this expression
	\begin{align}\label{giantK}
		K=\begin{pmatrix}
			t^1\bar{t}_1 &0 & 0 & t^1\bar{t}_4 \\
			0 & 0 & 0 & 0 \\
			0 & 0 & 0 & 0 \\
			t^4\bar{t}_1 &0 & 0 & t^4\bar{t}_4\\
		\end{pmatrix}.
	\end{align}
	Notice the notation of the $K$ matrix in our paper is  $|B\rangle=(\sum_{i,j}K_{ij}|ij\rangle)^L$, which is different from the notation in \cite{Gombor:2021hmj}.
	
	To check the integrable property of the giant graviton, we can utilize a special Bethe root that does not satisfy the root pairing and hence leads to a vanishing overlap. This requirement serves as a necessary condition for integrability. Here we consider the Bethe root
	\begin{align}
		&(L,K_u,K_v,K_w)=(3,1,2,1)\\\nonumber&\{u_1= 0.866025,w_1= 0.866025,v_1= -0.198072,v_2= 0.631084\},
	\end{align}
	which clearly does not satisfy the root pairing. Therefore, we expect the boundary state and the Bethe state to have a vanishing overlap. However, in order to ensure a non-zero overlap between the Bethe state and the equation \eqref{giantK}, it is necessary to adhere to the condition $K_u=K_v=K_w$. Fortunately, the $K$ matrix \eqref{giantK} can be rotated to achieve a non-vanishing overlap with the unpaired Bethe solutions.
	
	Let us first review the integrable boundary state through the boundary Yang-Baxter equation. An integrable boundary state will solve this equation
	\begin{align}\label{BYBE}
		R_{12}(u-v) K_{1}(u) R_{12}(u+v) K_{2}(v)=K_{2}(v) R_{12}(u+v) K_{1}(u) R_{12}(u-v).
	\end{align}
	Indeed, the integrable property remains conserved under an $SU(4)$ rotation. This observation stems from the $SU(4)$ invariance of the Hamiltonian, where the resulting eigenvalues continue to adhere to the selection rules. In this context, we have specifically selected an $SO(4)$ subgroup.
	
	With this in mind, we can now proceed to construct a new $K$ matrix
	\begin{align}
		K_{new}=g K g^{-1}\quad g\in SO(4).
	\end{align}
	where
	\begin{align}
		g=e^{\theta J^{14}}e^{\theta J^{23}}e^{\theta J^{3,4}}
	\end{align}
	and $J^{ij}_{mn}=\delta^i_m\delta^j_n-\delta^i_n\delta^j_m$. Plugging into \eqref{giantgraviton}, we obtain 
	\begin{align}
		\mathcal{D}_{M|\mathcal{O}}\sim  \omega-\omega^2.
	\end{align}
	
	In the previous section, we provided a proof for the integrability of the maximal giant graviton that corresponds to $\omega=1$. For $\omega=0$, we obtain:
	\begin{align}
		\langle B_\theta|=\sum_{A_s,B_s=1,4}\langle A_1\bar{B}_1\cdots A_L\bar{B}_L|tr[t^{A_1}\bar{t}_{B_1}\cdots t^{A_L}\bar{t}_{B_L}],
	\end{align}
	with
	\begin{align}
		\begin{aligned}
			\mathrm{t}^{1} & =\left(\begin{array}{cc}
				e^{i \theta}  & 0 \\
				0 & -e^{-i \theta} 
			\end{array}\right), \quad \overline{\mathrm{t}}_{4}=\left(\begin{array}{cc}
				e^{i \theta}  & 0 \\
				0 & -e^{-i \theta} 
			\end{array}\right), \\
			\mathrm{t}^{4} & =\left(\begin{array}{cc}
				-e^{i \theta}  & 0 \\
				0 & -e^{-i \theta} 
			\end{array}\right), \quad \overline{\mathrm{t}}_{1}=\left(\begin{array}{cc}
				e^{i \theta}  & 0\\
				0 & e^{-i \theta} 
			\end{array}\right).
		\end{aligned}
	\end{align}
	The argument presented in the previous section directly applies to this case, and thus we can conclude that this boundary state is also integrable. However, on the string theory side, $\omega=0$ corresponds to a giant graviton \cite{Yang:2021kot}, whereas its interpretation on the gauge theory side remains uncertain.
	
	To summarize, our analysis reveals that there are only two integrable cases for the giant graviton: $\omega=0$, which results in the degeneration of the three-point functions, and $\omega=1$, which corresponds to the maximal giant graviton. Notably, both of these cases are found to be integrable via our methodology.
	
	Notice that this conclusion contradicts the conjecture in \cite{Yang:2021hrl}, which proposes that non-maximal giant gravitons are also integrable. However, in \cite{Yang:2021hrl}, it was shown that the Bethe roots that do not satisfy the root pairing also fail to satisfy the zero momentum condition. As a result, the shift two symmetry of the boundary state results in a vanishing overlap between the boundary state and these Bethe roots. Hence, the calculations in this section imply that only two giant gravitons correspond to the integrable boundary state.
	\section{Conclusion}
	This paper demonstrates that the matrix product state proposed in \cite{Yang:2021hrl} is integrable, and that the integrable boundary condition naturally leads to root pairing. To achieve this, we divide the entire spin chain into the invariant part (PK factor) and their interval, using the indices of the transfer matrix. The final state is then determined by the Kronecker delta function involving the index $A$. This technique can be directly applied to the matrix product state consisting solely of commuting matrices.
	
	As an application, we utilize the fact that the Bethe root, which does not satisfy the root pairing, must have a vanishing overlap with the integrable boundary state, and the integrable property is preserved under a $SO(4)$ rotation. We employ these insights to prove that the matrix product state has only two integrable cases: $\omega=0$ or $1$.
	\subsection*{Acknowledgement}
	The author thanks Yunfeng Jiang for suggesting this topic. We also thank Yunfeng Jiang and Junbao Wu for useful discussions. We thank Tamas Gombor and Yinan Wang for pointing out the mistakes in the draft. This work is partly supported  by the National Natural Science Foundation of China, Grant No.~11975164, 11935009, 
	and  Natural Science Foundation of Tianjin under Grant No.~20JCYBJC00910

	\appendix
	\section{An Example}
	
	Let us consider an example to illustrate the proof idea. Taking L=3, we aim to demonstrate
	\begin{align}\label{example}
		&\sum_{\bar{B}_i,B_i=1,4}\mathrm{tr}[((u I+P)(-(u+2)I+K ))^3] |B_1\bar{B}_1B_2\bar{B}_2B_3\bar{B}_3\rangle=\\\nonumber &\sum_{\bar{B}_i,B_i=1,4} \mathrm{tr}[((-(u+2)I+P)(u I+K))^3] |B_1\bar{B}_1B_2\bar{B}_2B_3\bar{B}_3\rangle.
	\end{align}
	
	The expansion of the transfer matrices can be categorized into two classes:
	
	1. Terms that solely consist of the $PI, IK, II$ factors. For instance, $\mathrm{tr}(IIIIII)$, $\mathrm{tr}(IIPIPI),$ $\mathrm{tr}(IIIKIK)$. In these cases, we can establish a correspondence between the LHS and the RHS
	\begin{align}
		\mathrm{tr}(IIIIII)\leftrightarrow \mathrm{tr}(IIIIII), \mathrm{tr}(IIPIII)\leftrightarrow \mathrm{tr}(IIIKII),
		\mathrm{tr}(IIPIPI)\leftrightarrow \mathrm{tr}(IIIKIK)\cdots
	\end{align}
	A single $PI/IK$ factor in the trace corresponds to $\delta_{A_i B_i}$, which represents the index of the identity operator. By examining the coefficients, we can confirm an exact match. As discussed previously, the presence of $PI\cdots PI/IK\cdots IK$ induces a permutation among the indices on the odd/even sites. Therefore, we can deduce that
	\begin{align}
		\mathrm{tr}(IIPIPI)|\Psi\rangle= \mathrm{tr}(IIIKIK)|\Psi\rangle, \cdots
	\end{align}
	For example, one can check
	\begin{align}
		&\mathrm{tr}(PIIIPI)|\Psi\rangle=\sum_{B_i,\bar{B}_i=1,4}|B_3\bar{B}_1B_2\bar{B}_2B_1\bar{B}_3\rangle=\\\nonumber&\mathrm{tr}(IKIIIK)|\Psi\rangle=\sum_{B_i,\bar{B}_i=1,4}|B_1\bar{B}_3B_2\bar{B}_2B_3\bar{B}_1\rangle
	\end{align}
	
	2. For terms that include the $PK$ factor, the transfer matrix can be further classified into the following cases:
	
	Case 1: The term solely comprises $PK$ factors, such as $\mathrm{tr}(IIPKPK)$,$ \mathrm{tr}(IKPKPI),$ $\mathrm{tr}(IIIKPI),$ and so on. It is worth noting that the boundary state exhibits a two-site translation invariance, resulting in
	\begin{align}
		\mathrm{tr}(IIIKPI)|\Psi\rangle=\mathrm{tr}(PIIIIK)|\Psi\rangle.
	\end{align}
	
	It is necessary for the $PK$ factor to have the same position in both the LHS and the RHS. Then we establish a correspondence between the terms on the LHS and RHS
	\begin{align}
		\mathrm{tr}(PIIKII)\leftrightarrow \mathrm{tr}(PIIKII), \mathrm{tr}(IKIIPI)\leftrightarrow \mathrm{tr}(IKIIPI),\cdots
	\end{align}
	It is evident that there exists a correspondence between both sides in this case.
	\begin{align}
		\mathrm{tr}(PIIKII)|\Psi\rangle=\mathrm{tr}(PIIKII)|\Psi\rangle.
	\end{align}
	
	Case 2: When the interval between the PK factor contains $IK$ and $PI$ factors, we establish the correspondence between the terms
	\begin{align}
		&&\mathrm{tr}(\cdots\underbrace{{\color{red}PI\cdots IK}}_{\text{PK factor}}\underbrace{II\cdots}_{\text{only I}} \underbrace{IK\cdots IK\cdots}_{\text{number of K}=m} \underbrace{II\cdots}_{\text{only I}}\underbrace{PI\cdots PI\cdots}_{\text{number of P}=n} \underbrace{II\cdots}_{\text{only I}}\underbrace{ {\color{magenta}PI\cdots IK}}_{\text{PK factor}}\cdots)|\Psi\rangle\leftrightarrow\\\nonumber &&\mathrm{tr}(\cdots\underbrace{{\color{red}PI\cdots IK}}_{\text{PK factor}} \underbrace{II\cdots}_{\text{only I}}\underbrace{IK\cdots IK\cdots}_{\text{number of K}=n}\underbrace{II\cdots}_{\text{only I}} \underbrace{PI\cdots PI\cdots}_{\text{number of P}=m} \underbrace{II\cdots}_{\text{only I}}\underbrace{{\color{magenta}PI\cdots IK}}_{\text{PK factor}}\cdots)|\Psi\rangle.
	\end{align}
	
	Setting L=3, we have the correspondence
	\begin{align}
		\mathrm{tr}(IKPKIK)\leftrightarrow \mathrm{tr}(PIPKPI), \mathrm{tr}(IKIKPI)\leftrightarrow \mathrm{tr}(IKPIPI)\cdots
	\end{align}
	Acting on the boundary state, we obtain
	\begin{align}
		&\mathrm{tr}(IKPKIK)|\Psi\rangle=2\sum_{A,B_i,\bar{B}_i=1,4}|B_1\bar{B}_2AAB_2\bar{B}_1\rangle=\\\nonumber&\mathrm{tr}(PIPKPI)|\Psi\rangle=2\sum_{A,B_i,\bar{B}_i=1,4}|B_2\bar{B}_1AAB_1\bar{B}_2\rangle,\\\nonumber
		&\mathrm{tr}(IKIKPI)|\Psi\rangle=2\sum_{A,B_i,\bar{B}_i=1,4}|B_1AB_2\bar{B}_2A\bar{B}_2\rangle=\\\nonumber&\mathrm{tr}(IKPIPI)|\Psi\rangle=2\sum_{A,B_i,\bar{B}_i=1,4}|B_1AB_2\bar{B}_2A\bar{B}_2\rangle,\cdots
	\end{align}
	One can directly check these correspondence.
	
\end{document}